\documentclass[aps,twocolumn,float,prd,psfig]{revtex4}

\input epsf
\usepackage{graphicx}
\newcommand{\beq}{\begin{equation}}
\newcommand{\beqa}{\begin{eqnarray}}
		  \newcommand{\eeq}{\end{equation}}
\newcommand{\eeqa}{\end{eqnarray}}

\newcommand{\lsim}{\lesssim}
\newcommand{\gsim}{\gtrsim}

\newcommand{\vect}[1]{\mbox{\boldmath${#1}$}}
\newcommand{\lmk}{\left(}
\newcommand{\rmk}{\right)}
\newcommand{\lnk}{\left\{ }
\newcommand{\rnk}{\right\} }

\newcommand{\p}{\partial}

\newcommand{\vex}{{\vect x}}

\newcommand{\ven}{\vect n}

\newcommand{\ves}{{\vect s}}
\newcommand{\vev}{{\vect v}}

\begin{document}

\title{ Relativistic Astrophysics  with  Resonant Multiple Inspirals} 
%
%

\author{Naoki Seto$^{1}$, Takayuki Muto$^{1,2}$}
\affiliation{$^{1}$Department of Physics, Kyoto University
Kyoto 606-8502, Japan\\
$^{2}$Department of Earth and Planetary Sciences, Tokyo Institute of 
Technology, 2-12-1, Oh-okayama, Meguro-ku, Tokyo, 152-8550, Japan
}

\date{\today}

%
%
%
%
%
\begin{abstract}

We show that a massive black hole binary might resonantly trap a small third body ({\it e.g.} a neutron star) down to a stage when the binary becomes relativistic due to its orbital decay by gravitational radiation.  The final fate of the third body would be quite interesting for relativistic  astrophysics.  For example,  the parent binary could expel the third body with a velocity more than 10 \% of the speed of light.  We also discuss the implications of this three-body system  for direct gravitational wave observation.

\end{abstract}
\pacs{PACS number(s): 95.85.Sz 95.30.Sf}
\maketitle

\section{Introduction}
The orbital period of Pluto is 3/2 times that of Neptune, and their
mutual stability is maintained by  this simple commensurability (termed
{\it the mean motion resonance}).  In addition to this well-known fact,
the Solar system has various forms of the  mean motion resonances
\cite{ssd}.  Furthermore, more than 8 exoplanet systems are known to
have two planets in mean motion resonances.  One of  the notable
properties here is that once two planets are trapped in a stable
resonance relation, they often keep the state for a long time.  For
example,  recent numerical studies showed that some resonant trappings
are strong enough to be preserved against planet migration during which
orbital radii of two planets  decreased by more than 1 order of
magnitude \cite{Lee:2008nv}. 

Black hole (BH) binaries are considered as fascinating sources of broad
astrophysical phenomena, and, at the same time, their inspirals and
mergers are promising targets for gravitational wave (GW) observation
that would also provide us with ideal opportunities to test fundamental
aspects of gravity. As in the case of planetary systems, a massive BH
binary  might resonantly trap a small body,
 {\it e.g.} through
interaction with its surrounding disk, and  shrink its orbit by
dissipative processes. In this paper we call this kind of three-body
system  a resonant multiple inspiral (RMI), and study 
 the evolution and the
final fate of an RMI, including the post-Newtonian effects. We find that a BH binary has potential to keep a
third body down to a stage when the binary becomes relativistic due to
its orbital decay by gravitational radiation.  We also mention impacts
of RMIs on relativistic astrophysics and future GW observation.

\section{Stable Equilibrium Points}
We study the evolution of RMI  using
 a circular restricted three-body problem, namely,
analyze  the
motion of a test particle trapped by a BH binary in a circular
orbit with masses $(1-\mu)M$ and $\mu M$  [$M$: total mass of the
binary, $\mu(\le 0.5)$: the mass ratio]. We use the geometrical unit
$G=c=M=1$ with which the Schwarzschild radii of the two BHs are
$2(1-\mu)$ and $2\mu$ respectively. If necessary, we intentionally
recover the mass parameter $M$ to show the actual scales of physical
quantities. We introduce  a parameter $r_b$ for the binary separation. 

In this paper, we concentrate on  1:1 mean motion resonances as
tractable but intriguing examples  (see {\it e.g.} \cite{Wardell:2002iq}
for recent related studies).   In this case  the test particle moves
around the equilibrium points $L_4$ or $L_5$.  The two points are at
almost equilateral positions relative to the parent binary on its
orbital plane. After the pioneering work by Lagrange in 1772, various
properties  have been theoretically investigated  for the two points.  
In the real Universe, the Sun-Jupiter system ($\mu \sim 10^{-3}$) has a
large number of asteroids known as Trojans around its $L_4$ and $L_5$
(first discovered in 1906) \cite{ssd,tr1}, and their origin is still on
active debates \cite{tr1,tr2}. Similar objects have been found {\it
e.g.} for 
the Sun-Mars ($\mu\sim 2\times 10^{-7}$) or 
Saturn-Tethys
($\mu\sim 10^{-6}$) systems \cite{ssd}. Here Tethys is the fifth largest
moon of Saturn.

In order to describe
 the
 motion of a test particle around $L_4$ or $L_5$,
it is convenient to introduce a normalized frame $(X_N,Y_N,Z_N)$ that is
co-rotating with   the parent binary around its barycenter. In this
frame, the larger BH is at $(\mu,0,0)$, the smaller one is at
$(-(1-\mu),0,0)$ and their separation is unity. The $Z_N$ axis is the rotation axis of the parent binary and  normal to its orbital
plane \footnote{With respect to  the coordinate system defined in Fig.3.1 of \cite{ssd}, we have the correspondences $X_n$:$-x$, $Y_N$:$y$ and $Z_N$:$z$.}. Meanwhile, the positions of the two equilibrium points $L_4$ and
$L_5$ are given by $(X_L, -Y_L,0)$ and $(X_L, Y_L,0)$  with  
\beqa
X_L&=&-\frac12+\mu+\frac{5(-1/2+\mu)}{4r_b}+o(r_b^{-1}),\label{xl}\\
Y_L&=&\frac{\sqrt{3}}2+\frac{6\mu(1-\mu)-5}{8\sqrt{3}r_b}+o(r_b^{-1}),\label{yl}
\eeqa
where the terms $\propto r_b^{-1}$ represent the first-order
post-Newtonian (1PN) corrections \cite{DP,krefetz}. 

We briefly summarize the 
results of the
 linear stability analysis for a test particle
around the equilibrium points $L_4$ and $L_5$ \cite{gascheau, ssd}.  In
Newtonian mechanics,  
 small perturbations  around these points are given by 
the
superposition of two
stable oscillating  modes for $\mu<\mu_1=1/2-\sqrt{69}/18=0.038521$
\cite{gascheau}. They are known as the epicyclic motion with the
frequency $\omega_E$ and the libration motion with $\omega_L(<\omega_E)$
\cite{ssd}.  
The
 two frequencies are given by  
\beq
\omega_{E,L}=\omega_{BN} \frac{\sqrt{1\pm \sqrt{1-27\mu(1-\mu)}}}{\sqrt{2}}
\eeq
with the orbital frequency of the parent binary 
$\omega_{BN}\equiv(M/r_b^3)^{1/2}$  defined at the Newtonian order
(correspondence of signs, $\omega_E:+$ and $\omega_L:-$) \cite{ssd}. The
two frequencies  degenerate at $\mu=\mu_1$. For a larger mass ratio
$\mu>\mu_1$, the perturbation becomes unstable.

  With 1PN analysis,  the two basic frequencies $\omega_E$ and $\omega_L$ have
the  correction terms of $O(r_b^{-1})$. It is worth stressing   that the
ratio  $\omega_E/\omega_L$ now depends not only on the mass ratio $\mu$
but also on the separation $r_b$ of the parent binary.  The critical
mass ratio $\mu_1$  is given explicitly by $\mu_1=0.03852-0.29056 M/r_b$
\cite{DP}.


\section{Evolution of RMI}
An interesting question here is how the resonant trapping
changes with the evolution of the parent binary. Below, we numerically
study this issue, as a  restricted circular three-body problem. 

For the evolution of the circular  parent BH 
 binary, we
 can follow its 1PN orbit almost analytically,  including the orbital decay by the radiation of GW (see the Appendix for formulations and basic numerical scheme). 
Gravitational radiation reaction generates a dissipative force starting
at 2.5PN order \cite{LL1}.  Below this order, the system is
conservative. 
Since we assume  that the mass of 
the third body (test particle) is
negligible,  the dissipative
evolution is controlled by the motion of  the parent binary, through the fifth time derivative of its quadrupole moment.
After some algebra, the decrease of the orbital separation of the parent binary 
is described by the
standard expression \cite{LL1} 
\beq
\frac{dr_b}{dt}=-\frac{64}5 \mu (1-\mu) \lmk\frac{r_b}{M}  \rmk^{-3}.
\eeq
Accordingly, the rotation cycle $N$ of the binary before the coalescence
is estimated by 
\beq
N=\frac{1}{64\pi \mu (1-\mu) }\lmk\frac{r_b}{M}  \rmk^{5/2}.
\eeq
Given the positions of the parent binary, we  can numerically follow the evolution of the test particle.
Note that the dissipative acceleration directly works also on the test particle,
as easily understood from the fact that the energy loss rate is
proportional to the square of the GW amplitude \cite{LL1}.  This effect
might induce interesting effects for a 1:1 resonance, due to the apparent phase
coherence of the three particles.

Since we cannot properly handle the strong gravity around the 
two BHs with our
1PN equation of motion, we conservatively terminate our integration when
(i) the separation of the parent BH binary reaches the
innermost stable circular orbit  at $r_b=6$ \cite{LL1} or (ii)
similarly,  the test particle goes into 3 Schwarzschild radii of each
parent BH.  As for the initial position and velocity of the test
particle, we simply introduce two small parameters $q_x$ and $q_z$
related to its position in the corotating frame $(X_{N},Y_{N},Z_{N})$ as  
$
(X_L(1+q_x),\pm Y_L, 2X_Lq_z)
$
with $X_L$ and $Y_L$ given in eqs.(\ref{xl}) and (\ref{yl}) at the 1PN order. 
Here the parameter $q_x$ sets the perturbation of the third body  within the orbital plane of the parent binary,
and another one $q_z$ is related to the inclination with respect to the plane.  The particle is released  with an initial velocity with which
it is at rest in  the corotating frame. Note that we made perturbative
treatment of the 1PN correction terms. Thus, even with $q_x=q_z=0$, a
small initial perturbation of $O(r_b^{-2})$ is induced  around the
equilibrium points due to the truncation effect  (see {\it e.g.} \cite{Lousto:2007ji} for a similar
case). 

We hereafter discuss the results only from the $L_5$ region.
We have compared the evolution of the test particles
starting from both the $L_4$ and $L_5$ regions, but have not found 
notable qualitative differences. 
In Fig.1 we show the orbits of a test particle
at three  different epochs.  This is  a typical example for the
evolutionary behavior of the perturbation around the $L_5$ point. Each
figure clearly shows the characteristic profile of a tadpole orbit that
is given by 
the
superposition of the two elliptical motions; the
short-period  epicyclic mode (with the frequency $\omega_E$) and the
long-period libration mode (with $\omega_L$) \cite{ssd}.

\begin{figure}
  \begin{center}
\epsfxsize=8.cm
\begin{minipage}{\epsfxsize} \epsffile{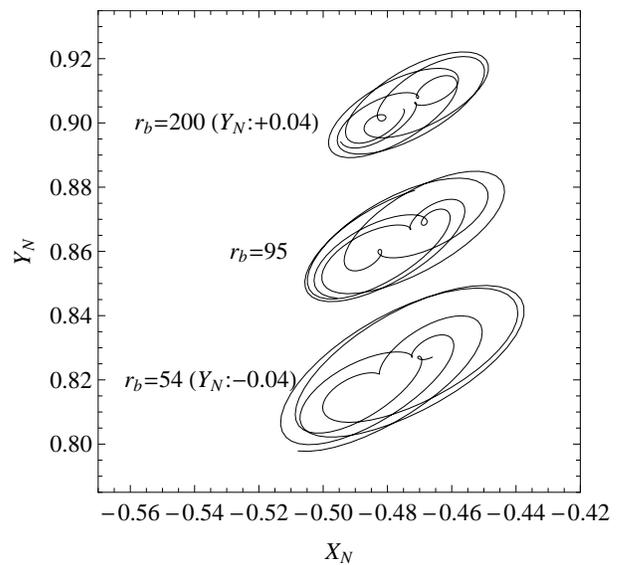}\end{minipage}
 \end{center}
  \caption{Evolution of the  orbit of a test particle around the $L_5$ point in the normalized corotating frame $(X_N,Y_N)$.  The mass ratio of the parent binary is $\mu=0.027$.   We show the orbits   at three  different separations $r_b=200$, 95 and 54. Each figure is given  for $\sim 8$ rotation cycles of the parent binary.  The initial conditions are  $q_x=0.002$, $q_z=0.001$ at  $r_{b}=200$. The upper and  bottom figures are shifted toward the vertical direction by $\pm 0.004$.
 }
\end{figure}

With  a simplified Hamiltonian and an associated adiabatic invariant \cite{LL2},
 Fleming  and Hamilton \cite{Fleming:2000qg} analytically
predicted how  migration of Jupiter affects the orbital evolution of its
Trojan asteroids (with Newtonian mechanics). Since the time scale of
gravitational radiation is much larger than the orbital period of the
parent BH binary  (at least for $r_b$  given in Fig.1), their analytical
prediction might fit  with our study.  Indeed, the evolution shown in
Fig.1 are close to the predicted scaling behavior $\propto r_b^{-1/4}$
for  the trajectories in the normalized corotating frame.  Thus we can
expect a fairly moderate evolution  for  the perturbation around the
$L_4$ and $L_5$ points during the inspiral of the circular parent binary, before
its separation becomes small and the unstable modes are generated due to
the PN effect
 as we see next.

\begin{figure}[!t]
  \begin{center}
\epsfxsize=8.cm
\begin{minipage}{\epsfxsize} \epsffile{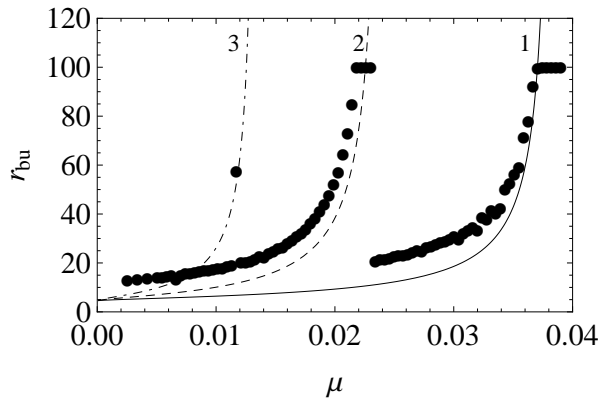} \end{minipage}
 \end{center}
  \caption{The unstable separation $r_{bu}$ of a parent binary as a
 function of its mass ratio $\mu$. The points are given from the
 evolution of the test particle  initially placed around the $L_5$ point  at
 $r_{b}=100$.  The solid curve is  derived from the stability condition
 $\omega_L=\omega_E$ at 1PN and  has the asymptotic profile $r_{bu}\to
 \infty$ at 
$\mu = {0.038521}$.
The dashed and dot dashed curves are given  for $\omega_E=2\omega_L$ and
 $\omega_E=3\omega_L$ with the critical mass ratios 
$\mu = {0.024293}$
and 
$\mu = {0.013516}$.
 respectively. 
 }
\end{figure}

For various mass ratio $\mu$, we followed a test particle initially with
$q_x=q_z=0$ at $r_b=100$, and measured the binary separation $r_{bu}$
where  the distance between the test particle and the $L_5$ point
becomes larger than 0.7 (a somewhat arbitrary value) in the normalized
corotating frame. In Fig.2, our numerical results are shown by
black
points. In addition to these results, we examined the unstable
separation $r_{bu}$ starting from finite (but small) $q_x$ and $q_z$,
and obtained similar results (see {\it e.g.} Figs.3 and 4). 
We also studied the evolution by Newtonian equation of motion, and found
that the binary can resonantly hold the test particle typically down to our
termination limit at
$r_b=6$.  In Fig.2,  the solid curve represents the
1PN prediction for the binary separation $r_{bu}$ where the $L_4$ and
$L_5$ points become unstable (corresponding to the degeneracy condition
$\omega_E=\omega_L$).  At the regime $\mu \gsim 0.025$, the unstable
radius $r_{bu}$ shows a reasonable agreement with the analytical result.

Interestingly, our numerical results in Fig.2 show a strong branch
around $\mu\sim 0.02$.  We found that this is caused by a resonant
instability due to the coupling of the epicyclic and libration modes
around $\omega_E=2\omega_L$, corresponding to the specific mass ratio
$\mu=0.024293$ for the Newtonian analysis \cite{res}. With the PN
effects, the ratio $\omega_E/\omega_L$ depends not only on the mass
ratio $\mu$ but also on the binary separation $r_b$, as mentioned
before.  Therefore, even if a parent  binary has a mass ratio satisfying
$\omega_E/\omega_L>2$ at the Newtonian limit $r_b\to \infty$, it could
match the unstable condition $\omega_E/\omega_L=2$ at a finite $r_b$ due
to the orbital shrink by  gravitational radiation.  As a result, with
the general relativistic effects, the binary is affected by the resonant
instability for a broader mass range $\mu$, 
in contrast
with a purely
Newtonian system.

Figure 2 also shows a branch associated with $\omega_E=3\omega_L$
corresponding to the specific mass ratio $\mu=0.013516$ for the
Newtonian limit \cite{res}. 
However,
 this higher-order effect is relatively
weak, and many binaries can safely go through the unstable separation,
as shown in Fig.2.

\section{Final Fate of RMI}
As we 
have
studied so far, with RMI, a third body could stay around a parent
BH binary deeply into the relativistic regime.  Even though  careful
attention should be paid for interpreting our 1PN results, they would
provide us with  qualitative insights about the potential  final fate of the third
body.  This issue would be particularly interesting in relation to GW
observation with the Laser Interferometer Space Antenna (LISA)
\cite{lisa}.  In order to make concrete pictures for LISA, we assume
that the total mass of a parent BH binary is $M\sim 10^6M_\odot$, and
the mass of 
the
third body is $\sim 1-10M_\odot$. 
Given the stability condition $\mu<\mu_1=0.03852$ for a 1:1 resonance,
the parent BH binary  might be regarded as an intermediate mass ratio
inspiral  rather than an inspiral of two comparable BHs.  While
abundance of BHs around $\sim 10^4M_\odot$ is not well known at present,
the actual event rate of the former might be higher than that of the
latter.

The signal-to-noise ratio of GW associated with the small third body
would be much lower than that of the parent BH binary. 
However,
the stronger
GW signal from the parent BH binary would enable us to easily estimate
the basic parameters of the parent such as the mass ratio $\mu$ or the
total mass $M$.  Then, for example, from the estimated masses of a
potential parent, we can predict the epoch when the $L_4$ and $L_5$
points become dynamically  unstable (see Fig.2).  This kind of  prior
information would considerably help us to make a  careful follow-up data
analysis searching for a weaker  GW signal by a third body.

Statistical analyses for the final fates of  RMIs with realistic initial
conditions would be useful for astrophysical arguments, but they are far
beyond the scope of this paper. Here, we would rather make qualitative
discussions on the expected destinies.  Among our numerical samples,
ejection of a test particle from the parent binary is a frequent final
state. In Fig.3, we show the trajectory of the test particle evolved
from Fig.1.  The $L_5$ point becomes dynamically  unstable around
$r_b\sim 24.6$ and the particle was soon scattered by the smaller BH at
$(-0.973,0,0)$. It escaped from the binary with a terminal velocity of
$\sim 0.14c$.
 which is much larger than the typical kick velocity by the
anisotropic GW emission \cite{Campanelli:2007cga}.   
For an RMI event,
the third body can be  scattered  by a relativistic binary, and this
magnitude is not surprising.  The third body should be a neutron star or
a black hole for surviving tidal disruption during the large angle
scattering by the smaller BH of the parent binary.

Though an electromagnetic wave  search for such a high velocity
object would be challenging, we might get a signature of its ejection by
careful analysis of GWs. 
If the third body is a white dwarf and tidal disruption occurs, we might
observe a short-period   electromagnetic wave signal before the merger of the parent BH
binary. This might help us to identify its host galaxy  and perform
cosmological studies, {\it e.g.} constraining dark energy parameters
through the relation between the redshift and the luminosity distance
of the binary 
\cite{Schutz:1986gp}.

From our numerical samples, we  expect that a plunge into the larger BH
would  be another likely scenario. 
Note that,  from a distance of
$O(r_{bu})$, its angular size  is much larger than a degree scale.  
For a
plunge into the smaller one, we might detect its GW signal by
ground-based detectors, depending on the mass of the BH.

\begin{figure}
  \begin{center}
\epsfxsize=7.cm
\begin{minipage}{\epsfxsize} \epsffile{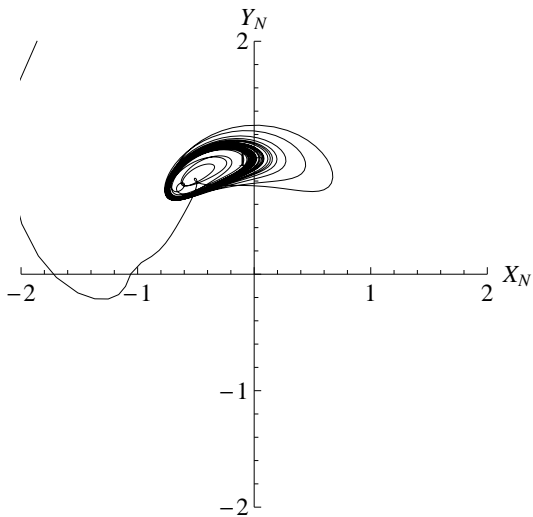}\end{minipage}
 \end{center}
  \caption{Ejection of a test particle  around the binary separation $r_b=24.6$ after $N\sim1.1\times 10^5$ cycles from $r_b=200$. The ejection velocity is $\sim 0.14c$.  The initial conditions are the same  as Fig.1.  The larger BH is at $(0.027,0,0)$ and the smaller one is at $(-0.973,0,0)$. They rotate counterclockwise.
 }
\end{figure}

\begin{figure}
  \begin{center}
\epsfxsize=7.cm
\begin{minipage}{\epsfxsize} \epsffile{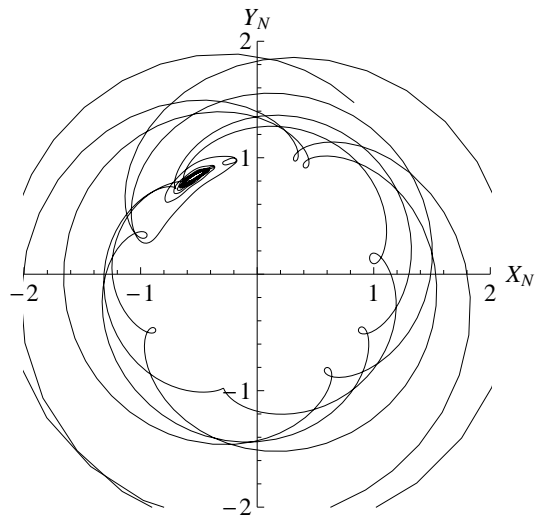}\end{minipage}
 \end{center}
  \caption{Dynamical formation of an EMRI system around  $r_b=13.0$ with $\mu=0.0025$. The initial positions are  $q_x=0.004$ and $q_z=0.08$ at $r_{b}=77$ ($9.9\times 10^4$ cycles to $r_b=13.0$).
 }
\end{figure}

Some of our numerical samples resulted in 
the
dynamical formation of
extreme-mass-ratio-inspiral (EMRI) systems.  In Fig.4, we present the
trajectory of a test particle around the unstable separation $r_b\sim
13$.  After the transitional stage shown in Fig.4, the test particle
almost decouples from the evolution of the parent BH binary.  When the
binary reach the innermost stable circular orbit $r_b=6$, we have an eccentric EMRI system with
pericenter distance $\sim 13$ and the apocenter distance $\sim 50$. 
GW from an EMRI is a very important target of LISA for directly probing highly distorted
geometry around BHs, although we need
significant effort to detect it
due to its complicated waveform and the limitation of available computational power \cite{Gair:2004iv}.
 As we commented
earlier, the basic parameters of the parent BH would be estimated by
its stronger but simpler GW signal.  Thus the subsequent EMRI signal
would be detected more easily than a blind EMRI search as usually
assumed. 
 Furthermore, we might, in principle, use the third body to
probe the dynamical gravitational field caused by the merger of the
parent BH binary  to precisely measure the 
basic parameters (e.g. mass and spin) of the merged BH.

\section{Discussions}

Our study for RMIs is based on the 1PN restricted three-body analysis
for a circular parent binary.  This simple treatment can be extended in
diverse ways. One of the principle directions is to include higher-order
PN effects and spins of BHs.  
A complementary 
approach would be to inject
a test particle into a numerically evolved BH binary (for recent
breakthrough see \cite{Pretorius:2005gq}).  
 From Newtonian
 analyses
we expect that modest eccentricity $(e\lsim
 0.1)$ of the parent binary would not largely change our basic results
 \cite{el}, but
this should be checked specifically.    Meanwhile, we
 have regarded the third body as a massless pointlike particle.
 Actually,  the assumption for the mass  would  be largely relaxed for
 RMIs \cite{gascheau}. For example, Saturn has stable co-orbital
 satellites (in a 1:1 resonance);  Janus and Epimetheus whose masses are
 comparable (1:0.25) \cite{ssd}.  Furthermore, the internal structure of
 the third body might cause interesting astrophysical effects ({\it
 e.g.} for  a main-sequence star or even a gas cloud). 
Beyond a three-body problem,   a parent BH binary might keep many small objects around
 its $L_4$ and $L_5$ points, similar to the Trojan asteroids of the
 Sun-Jupiter system.   
In the case of a 1:1  resonance, we have a severe upper limit
 $\mu<\mu_1$ for the restricted three-body problem, but
a larger mass
 ratio might be allowed for resonances other than 1:1.

In addition to these refinements for  the
evolution of RMIs, follow-up
studies in related fields are worth exploring. 
More detailed analyses
for detectability of GW signatures at various
stages of RMIs would be fruitful, especially for space interferometers
such as LISA.  Finally,   discussions on the formation mechanism of RMIs
would be an attractive topic on astrophysics (see {\it e.g.} \cite{Fujii:2010rd} and references there in for related studies).

The authors would like to thank  H. Arakida and K. Tomida  for their
kind support. 
This work was supported by the Grant-in-Aid for the Global COE Program
"The Next Generation of Physics, Spun from Universality and Emergence"
from the Ministry of Education, Culture, Sports, Science and Technology
(MEXT) of Japan. 





\appendix

\section{equations of motion for  particle systems}
In this Appendix, we briefly explain the basic equations applied  in   our numerical calculations for dynamical evolution of RMI.
For the conservative 1PN equations of motion, we use the Hamiltonian form of the Einstein-Infeld-Hoffman Lagrangian given for a  point-particle system with masses $m_a$ ($a$: the label for particles)  \cite{cc,LL1}.   We follow the positions $x_{ai}$ ($i=1,2,3$: the label for spatial directions)  in the barycentric non-rotating frame $(x_1,x_2,x_3)$.  Rather than adopting the standard conjugate momentums $p_{ai}$, we introduce the new variables
\beq
s_{ai}\equiv \frac{p_{ai}}{m_a}
\eeq
that are convenient for dealing with restricted three-body problems.  Our 1PN Hamiltonian is expanded as
\beq
H=H_N+H_{1PN}
\eeq 
with the explicit forms of the Newtonian and 1PN terms
\beqa
H_N&=&\frac12\sum_am_a s_a^2-\frac12\sum_{a,b\ne a} \frac{m_a m_b}{r_{ab}}\\
H_{1PN}&=&-\frac18\sum_a m_a (s_a^2)^2+\frac12\sum_{a,b\ne a,c\ne a}\frac{m_a m_b m_c}{r_{ab} r_{ac}},\nonumber\\
& & +\frac14 \sum_{a,b\ne a} \frac{m_a m_b}{r_{ab}} \{ -6 {s_a^2}+7 \ves_a\cdot\ves_b \nonumber\\
& &~~~~~~~~~+(\ven_{ab}\cdot \ves_a)(\ven_{ab}\cdot \ves_b) \},
\eeqa
where $r_{ab}=|\vex_a-\vex_b|$ and $\ven_{ab}=(\vex_a-\vex_b)/r_{ab}$.

For our new set of variables $(x_{ai},s_{ai})$, the canonical equations are modified  as follows;
\beq
\frac{ds_{ai}}{dt}=-\frac1{m_a}\frac{\p H}{\p x_{ai}},~~~\frac{dx_{ai}}{dt}=\frac1{m_a}\frac{\p H}{\p s_{ai}}\label{mh}.
\eeq
These are well behaved in the limit $m_a\to 0$  for a restricted problem.
Note that with the 1PN 
terms
the simple identity
$s_{ai}= \frac{dx_{ai}}{dt} (\equiv v_{ai})$ does not hold, and we have the following correspondence at the 1PN order \cite{cc};
\beqa
s_{ai}&=&v_{ai}+v_{ai}\lmk \frac{v_a^2}2+3 \sum_{b\ne a}\frac{ m_b}{r_{ab}}  \rmk\nonumber\\& &-\frac12\sum_{b\ne a} \frac{m_b}{r_{ab}}\lnk 7v_{bi}+(\ven_{ab}\cdot \vev_b)n_{abi}  \rnk.
\eeqa

In this paper, we study circular restricted three-body problems in which the massless third body is irrelevant for the evolution of the parent binary (with masses $m_1$ and $m_2)$.  The dynamics of the circular parent binary is characterized by the orbital frequency $\omega_b$ as a function of the binary separation $r_b\equiv |\vex_1-\vex_2|$, and it is perturbatively expressed as 
\beq
\omega_B=\omega_{BN}+\omega_{B1PN}
\eeq
with the Newtonian result
\beq
\omega_{BN}=\lnk \frac{m_1+m_2}{r_b^3}\rnk^{1/2}.
\eeq
Applying the above Hamiltonian for two particles, we  obtain the 1PN correction term as (see {\it e.g.} \cite{krefetz})
\beq
\omega_{B1PN}=\omega_{BN} \frac{m_1+m_2}{2r_b} \lnk \frac{m_1 m_2}{(m_1+m_2)^2}-3  \rnk.
\eeq
Thus the motion of the parent binary can be handled analytically, and we can numerically  follow the position of the third body with  the modified canonical equations,  plugging in the analytical information of the parent binary.


Up to now, our 1PN system is conservative and this will provide us with a good opportunity to check the accuracy of our numerical code (based on a fifth order Runge-Kutta integration scheme \cite{nr})  for the motions of  third bodies around the Lagrange points.  For relevant sets of parameters $(r_b,\mu,q_x)$, we examined the 1PN version of Jacobi energy derived for motions of the third body within the orbital plane of the parent binary \cite{md}. We found that it is typically conserved by $\sim10^{-5}$ level for $\sim 10^5$ orbital rotation cycles.

Finally we summarize the dissipative effects caused by gravitational radiation. The first equation of (\ref{mh}) is now modified as follows 
\beq
\frac{ds_{ai}}{dt}=-\frac1{m_a}\frac{\p H}{\p x_{ai}} +\lmk \frac{ds_{ai}}{dt}\rmk_{rad},
\eeq
where the second term is due to the radiation, and at the lowest order, it is given by
\beq
\lmk \frac{ds_{ai}}{dt}\rmk_{rad} =-\frac2{15} D_{ij}^{(V)} x_{aj}.\label{ds}
\eeq
Here $D_{ij}^{(V)}$ is the fifth time derivative of the quadrupole moment $D_{ij}$  defined by \cite{LL1}
\beq
D_{ij}\equiv \sum_a m_a (3x_{ai} x_{aj}-x_a^2\delta_{ij}).
\eeq
For the restricted three body-problem, the moment $D_{ij}$ is determined only by the parent circular binary ($a=1,2$), and its fifth time derivative is evaluated analytically with a standard adiabatic treatment.  After some algebra, we obtain the time derivative of the separation $r_b$ of the parent binary as in Eq.(4), and its orbital position is determined straightforwardly.  Then, with a method similar to the previous conservative case, we can numerically follow the massless third body.  Our numerical results in Sec.III-IV   are obtained in this manner.

\end{document}